\documentclass[aps,prl,floatfix,twocolumn,showpacs,10pt]{revtex4-2}

\usepackage{graphicx}
\usepackage{dcolumn}
\usepackage{bm}
\usepackage{amssymb}
\usepackage{amsmath} 
\usepackage{rotating} 
\usepackage{xcolor}

\begin{document}

\title{Quantum Many-body Theory from a Solution of the $N$-representability Problem}

\author{David A. Mazziotti}

\email{damazz@uchicago.edu}
\affiliation{Department of Chemistry and The James Franck Institute, The University of Chicago, Chicago, IL 60637}%

\date{Submitted September 15, 2022\textcolor{black}{; Revised December 13, 2022}\textcolor{black}{; Revised February 13, 2023}}


\begin{abstract}

Here we present a many-body theory based on a solution of the $N$-representability problem in which the ground-state two-particle reduced density matrix (2-RDM) is determined directly without the many-particle wave function.  We derive an equation that re-expresses physical constraints on higher-order RDMs to generate direct constraints on the 2-RDM, which are required for its derivation from an $N$-particle density matrix, known as $N$-representability conditions.  The approach produces a complete hierarchy of 2-RDM constraints that do not depend explicitly upon the higher RDMs or the wave function.  By using the two-particle part of a unitary decomposition of higher-order constraint matrices, we can solve the energy minimization by semidefinite programming in a form where the low-rank structure of these matrices can be potentially exploited.  We illustrate by computing the ground-state electronic energy and properties of the H$_{8}$ ring.

\end{abstract}


\maketitle

{\em Introduction:} The $N$-particle wave function or density matrix of a many-body system with at most pairwise interactions contains significantly more information than necessary to compute the ground-state energy and its one- and two-particle properties~\cite{Mazziotti2007, Coleman2000}.  Density functional theory (DFT) expresses the many-particle energy as a functional of the single-particle density, but its exact functional form for electrons as well as other particles remains unknown~\cite{DFTBook}, causing well-known problems with charge-transfer states, van der Waals interactions, band gaps, and more generally, static correlation~\cite{Medvedev2017}.  Another approach is to express the energy as a functional of the two-particle reduced density matrix (2-RDM)~\cite{Coleman1963}.  The 2-RDM arises from integration of the $N$-particle density matrix over all particles except two. Unlike the situation in DFT, the energy is a known, linear functional of the 2-RDM.  The energy expression, however, cannot be minimized with respect to the 2-RDM without additional constraints to ensure that the 2-RDM represents an $N$-particle system, which are known as $N$-representability conditions~\cite{Coleman1963, Garrod1964, Kummer1967, Erdahl1978, Erdahl1989, Coleman2000, Mazziotti2007, Mazziotti2012b}.  Minimization without these conditions yields energies significantly below the ground-state energy.  The search for the $N$-representability conditions is known as the $N$-representability problem~\cite{Coleman1963, Mazziotti2012b, Castillo2021}.

Kummer~\cite{Kummer1967} first proved the formal existence of a solution to the $N$-representability problem although his approach required solving a large number of $N$-particle eigenvalue equations.  More recently, the author~\cite{Mazziotti2012b} showed that the complete $N$-representability conditions are derivable from convex combinations of higher-particle, constraint matrices.  In the past decade approximate conditions have been employed in variational calculations of the 2-RDM~\cite{Erdahl2000, Nakata2001, Mazziotti2001c, Mazziotti2002b, Zhao2004, Mazziotti2004a, Mazziotti2005, Cances2006, Mazziotti2006b, Erdahl2007, Braams2007, Gidofalvi2008, Shenvi2010, Mazziotti2011, Verstichel2012, Baumgratz2012, Poelmans2015, Mazziotti2016, Alcoba2018, Rubio-Garcia2019, Head-Marsden2020, Haim2020, Han2020, Mazziotti2020, Li2021, Knight2022} to solve significant problems such as the elucidation of the electronic properties of a glassy, highly conductive amorphous polymer\textcolor{black}{~\cite{Xie2022}} and the revelation of exciton condensation in molecular analogs of double-layer graphene~\cite{Safaei2018, Schouten2021, Schouten2022, Sager2022} \textcolor{black}{(refer to Refs.~\cite{Schilling2021, Benavides-Riveros2020, Schilling2019, Piris2021, Piris2017a, Gibney2022a} for their use in one-particle RDM functional theories)}.  Nonetheless, despite these theoretical advances and applications, an open, critically important challenge is to include higher $N$-representability conditions on the 2-RDM to obtain a convergent series of increasingly tighter lower bounds on the energy.

In this Letter we present a many-body theory based on a solution of the $N$-representability problem in which the ground-state energy and 2-RDM are variationally computed without any explicit dependence on the higher-order RDMs or the $N$-particle wave function.  We derive an equation that parametrically generates a systematic, complete hierarchy of $N$-representability conditions on the 2-RDM from convex combinations of constraints on higher-order RDMs.  Using a unitary decomposition to extract the two-particle part of the higher-order constraint matrices~\cite{Coleman1980, AuChin1983}, we solve the energy minimization at each level of the hierarchy by an efficient optimization, known as semidefinite programming~\cite{Vandenberghe1996, Fukuda2007, Mazziotti2011, Erdahl1979}.  \textcolor{black}{Compared to earlier work~\cite{Mazziotti2001c, Mazziotti2006b, Gidofalvi2006, Schwerdtfeger2009, Li2021, Mazziotti2012b}, the present theory has a potentially major advantage in computational scaling because the constraint matrices, unlike the higher-order RDMs, are extremely low rank for Hamiltonians with pairwise interactions.}  While the theory is developed here for electrons which are fermions, with minor adjustments it is applicable to bosons~\cite{Gidofalvi2004}.  To illustrate, we apply the theory to compute the energies and 2-RDMs of the H$_{8}$ ring, which show rapid convergence with the level of the hierarchy.

{\em Theory:} Because electrons are indistinguishable with pairwise interactions, the electronic energy of any atom or molecule can be written in terms of only two electrons
\begin{equation}\label{eq:K2D2a}
  E = {\rm Tr} \left ( ^{2} {\hat K} \, ^{2} D(12;{\bar 1}{\bar 2}) \right )
\end{equation}
where
\begin{equation}\label{eq:K2}
  ^{2} {\hat K} = \frac{N}{2} \sum_{j=1}^{2}{ \left ( - \frac{{\hat \nabla}_{j}^{2}}{2} - \sum_{k}{\frac{Z_{k}}{r_{jk}}} \right ) } + \binom{N}{2} \frac{1}{r_{12}}
\end{equation}
is the two-electron reduced Hamiltonian operator and
\begin{equation} \label{e:D2}
^{2} D(12;{\bar 1}{\bar 2}) = \int{ \Psi(12 \ldots N) \Psi^{*}({\bar 1}{\bar 2} \ldots N) d3 \ldots dN }
\end{equation}
is the 2-RDM expressed in terms of the $N$-electron wave function $\Psi(12 \ldots N)$.  Each roman number corresponds to the spatial and spin coordinates of an electron.  In a matrix representation with respect to a finite set of orthonormal orbitals, the energy can be expressed as
\begin{equation} \label{eq:K2D1b}
E  =  \sum_{ijkl}{^{2} K^{ij}_{kl} \, ^{2} D^{ij}_{kl}}
\end{equation}
in which the indices $i$, $j$, $k$, $l$ range from 1 to $r$ with $r$ being the rank of the orbital basis set.

The 2-RDM in the energy expression must be further constrained to ensure that it represents an $N$-electron system---the $N$-representability conditions.   These conditions can be expressed by a system of linear inequalities on the 2-RDM
\begin{equation}\label{eq:B2D2}
  {\rm Tr} \left ( ^{2} B_{j} \, ^{2} D \right ) \ge 0
\end{equation}
in which the two-electron matrices $^{2} B_{j}$ lie in a convex cone $^{2} P^{*}_{N}$ that is polar to the set of 2-RDMs $^{2} P_{N}$.  The polar cone of $^{2} B_{j}$ matrices can be described by the following linear matrix inequalities (or semidefinite constraints)
\begin{equation} \label{eq:BN}
^{2} B_{j} \wedge I^{N-2} \succeq 0
\end{equation}
in which $\wedge$ is the Grassmann wedge product~\cite{Coleman1980,Slebodzinski1970, Mazziotti1998}, $I^{N-2}$ is the $(N-2)$-electron identity matrix, and $M \succeq 0$ indicates that the matrix $M$ is positive semidefinite (nonnegative eigenvalues).   Each $^{2} B_{j}$ defines the one- and two-body terms of a positive semidefinite $N$-electron matrix in Eq.~(\ref{eq:BN}) if and only if it has a nonnegative trace against every $N$-representable 2-RDM in Eq.~(\ref{eq:B2D2}).  Otherwise, there is a contradiction between the expectation values of $^{2} B_{j}$ on the reduced space and those on the $N$-electron space.  Each semidefinite constraint in Eq.~(\ref{eq:BN}), however, is not easily checked because it involves solving an $N$-electron eigenvalue equation.

The solution of the $N$-representability problem, as we showed previously, can be formally cast as convex combinations of higher-body positive semidefinite operators that generate the two-particle operators in the polar cone~\cite{Mazziotti2012b}.   Here we recognize that we can successfully express each convex combination by the following fundamental and practical matrix equation for parameterizing the $^{2} B_{j}$ matrices:
\begin{equation} \label{eq:B2}
\sum_{i}{^{p} B_{i,j}} - {}^{2} B_{j} \wedge I^{p-2} = 0
\end{equation}
where the $^{p} B_{i,j}$ matrices can be selected from known, necessary $N$-representability conditions on the $p$-RDM, called $p$-positivity conditions (see below), and hence, they have the property $^{p} B_{i,j} \wedge I^{N-p} \succeq 0$ by construction without the need to form the $N$-electron matrix.   Eq.~(\ref{eq:B2}) converts a sum of $p$-electron matrices that are necessary for a $p$-RDM to be $N$-representable into a 2-electron matrix $^{2} B_{j}$ that is necessary for a 2-RDM to be $N$-representable.  Formally, the constraints are necessary and sufficient when $p=N$ \textcolor{black}{because each member of the polar cone in Eq.~(\ref{eq:BN}) is generated}, but practically, based on previous $p$-RDM calculations~\cite{Mazziotti2001c, Mazziotti2006b, Gidofalvi2006, Schwerdtfeger2009, Sinitskiy2010, Li2021}, they often converge to either a complete or a practically complete set for $p \ll N$.  Therefore, the combination of Eqs.~(\ref{eq:B2D2}) and~(\ref{eq:B2}) provides a hierarchy of $N$-representability conditions for the variational computation of the 2-RDM.  \textcolor{black}{This hierarchy has a significant advantage over the hierarchy in Ref.~\cite{Mazziotti2012b} because it expresses all of the conditions through the parametrization of a single matrix equation.}

The sum over $i$ in Eq.~(\ref{eq:B2}) consists of $(p+1)$ terms that translate each of the $p$-positivity conditions on the $p$-RDM into conditions directly on the 2-RDM, which we call $(2,p)$-positivity conditions.  For 2-positivity there are three terms with the first term $^{2} B_{1,j} \succeq 0$ corresponding to the D2 condition in which the 2-particle RDM is constrained to be positive semidefinite.   Other contributions can be readily defined in second quantization~\cite{Coleman1963, Garrod1964, Kummer1967, Erdahl1978, Erdahl1989, Coleman2000, Mazziotti2007, Mazziotti2012b}
\begin{eqnarray} \label{eq:BQ}
\sum_{pqst}{^{2} B_{2,j}^{pq;st} {\hat a}^{\dagger}_{p} {\hat a}^{\dagger}_{q} {\hat a}^{}_{t} {\hat a}^{}_{s}} & = & \sum_{pqst}{^{2} B_{Q,j}^{st;pq} {\hat a}^{}_{s} {\hat a}^{}_{t} {\hat a}^{\dagger}_{q} {\hat a}^{\dagger}_{p}  } \\
\sum_{pqst}{^{2} B_{3,j}^{pq;st} {\hat a}^{\dagger}_{p} {\hat a}^{\dagger}_{q} {\hat a}^{}_{t} {\hat a}^{}_{s}} & = & \sum_{pqst}{^{2} B_{G,j}^{pt;sq} {\hat a}^{\dagger}_{p} {\hat a}^{}_{t} {\hat a}^{\dagger}_{q} {\hat a}_{s}  }
\end{eqnarray}
where $^{2} B_{Q,j} \succeq 0$ and $^{2} B_{G,j} \succeq 0$.  The $^{2} B_{2,j}$ and $^{2} B_{3,j}$ are obtained in terms of the positive semidefinite $^{2} B_{Q,j}$ and $^{2} B_{G,j}$ by rearranging the fermionic second-quantized operators on the right sides of the equations to match the canonical ordering on the left sides of the equations.   These terms correspond to the Q2 and G2 conditions that constrain the distributions of two holes and a particle-hole pair to be nonnegative.  Similarly, for the 3-positivity conditions there are four terms with the first term $^{3} B_{1,j} \succeq 0$ corresponding to the D3 condition in which the 3-particle RDM is constrained to be positive semidefinite.  Other terms correspond to the E3, F3, and Q3 matrices that constrain the probability distributions of two particles and a hole, two holes and a particle, and three holes~\cite{Mazziotti2001c, Mazziotti2006b, Mazziotti2012b}\textcolor{black}{; see Supplemental Material~\footnote{see Supplemental Material at URL for $^{3} B_{i,j}$ matrices of the 3-positivity conditions.}.}  Similarly, the convex combinations can be defined for the higher $p$-positivity conditions where $p > 3$~\cite{Mazziotti2001c} to generate a complete hierarchy of $N$-representability conditions on the 2-RDM, the $(2,p)$-positivity conditions~\cite{Mazziotti2012b}.

These $N$-representability conditions can be efficiently implemented for the direct variational calculation of the ground-state 2-RDM through the following semidefinite program based on the 2-RDM’s polar cone
\begin{align}
\max_{E,X} {E} ~~~~~~ & \\
{\rm such~that~~~} \left ( {}^{2} K - E \, I^{2} \right ) - {}^{2} B & = 0 \label{eq:c1} \\
\sum_{i}{{}^{p} B_{i}[X]} - {}^{2} B \wedge I^{p-2} & = 0 \label{eq:c2}
\end{align}
in which we cast the $^{2} B$ matrix, which is not necessarily positive semidefinite, as a functional of the ${}^{p} B_{i}[X]$ matrices, which are functionals of the positive semidefinite matrices $X$, via their unitary decompositions~\cite{Coleman1980, AuChin1983}
\begin{equation}
^{2} B = \alpha_{p,0} {\hat L}_{p}^{0}(^{p} B) I^{2} + \alpha_{p,1} {\hat L}_{p}^{1}(^{p} B) \wedge I^{1} + \alpha_{p,2} {\hat L}_{p}^{2}(^{p} B)
\end{equation}
where ${\hat L}_{p}^{q}$ is the contraction operator that contracts the matrix from a $p$-electron matrix to a $q$-electron matrix and
\begin{equation}
\alpha_{p,k} = \frac{  \left(-1\right)^{3-k} \left(r -p -3\right)! \left(r -p -2\right) p !^{2}}{\left(r -2-k \right)! \left(2-k \right)! \left(p -3\right)! \left(k -p \right) k !^{2}} .
\end{equation}
\textcolor{black}{This polar-cone formulation is critical to expressing the optimization as a semidefinite program without the higher RDMs.  Moreover, if the Hamiltonian has pairwise interactions and/or sparsity, it provides a framework for maximally exploiting the low-rank, sparse structure of the ${}^{p} B_{i}[X]$ and $X$ matrices.}

The unitary decomposition allows us to separate the solution of a given many-electron problem in Eq.~(\ref{eq:c1}) from the solution of the $N$-representability problem in Eq.~(\ref{eq:c2}).  Eq.~(\ref{eq:c1}) determines the $^{2} B$ matrix in the polar cone that generates the ground-state energy of any atom or molecule defined by its reduced Hamiltonian $^{2} K$ while Eq.~(\ref{eq:c2}) transforms the $N$-representability conditions of the $p$-RDM into a parameterization of the 2-RDM's polar cone.  While the 2-RDM does not appear explicitly in any of the above equations, in the Lagrange multiplier formulation of the polar-cone semidefinite program, the Lagrange multiplier of Eq.~(\ref{eq:c1}) is the 2-RDM.  This important result follows, as shown in detail in Ref.~\cite{Mazziotti2020}, from an application of the Hellmann-Feynman theorem~\cite{Feynman1939}.


\begin{table*}[t!]

\caption{The ground-state energy as well as several properties of the 1- and 2-RDMs for H$_{8}$ with $R=0.707$~\AA\ are shown from the 2-RDM method with different levels of $N$-representability conditions. \textcolor{black}{The trace of the 2-RDM is normalized to one.}}

\label{t:24}

\begin{ruledtabular}
\begin{tabular}{cddddddd}
& \multicolumn{2}{c}{Total Energy} & \multicolumn{2}{c}{Natural Orbitals} & \multicolumn{2}{c}{$\langle 1/r_{12} \rangle$} &  \multicolumn{1}{c}{\textcolor{black}{~~~2-RDM}} \\ \cline{2-3} \cline{4-5} \cline{6-7} \cline{8-8}
\text{Method} & \multicolumn{1}{c}{~~~~~Value} & \multicolumn{1}{c}{~~~~~Error} & \multicolumn{1}{c}{~~~~~HONO} & \multicolumn{1}{c}{~~~~~LUNO} & \multicolumn{1}{c}{~~~~~Value} & \multicolumn{1}{c}{~~~~~Error} & \multicolumn{1}{c}{~~~~~\textcolor{black}{Error}} \\ \hline
 \text{HF} & - 2.24549 &  0.11415 &  1.00000 &  0.00000 &  0.47040 &  0.00491 & \color{black}  0.053813 \color{black} \\
 \text{MP2} & - 2.30955 &  0.05009 & 0.99477 &  0.00733 &  0.46750 &  0.00200 & \color{black} 0.053149 \color{black} \\
 \text{CCSD} & - 2.34250 &  0.01714 &  0.65083 &  0.34956 &  0.46562 &  0.00012 & \color{black} 0.051282 \color{black} \\
 \text{2,2-POS} & - 2.39171 & - 0.03207 &  0.50075 &  0.50075 &  0.46354 & - 0.00195 & \color{black}  0.001523 \color{black} \\
 \text{2,2-POS+T1} & - 2.37721 & - 0.01757 &  0.50138 &  0.50138 &  0.46438 & - 0.00111 & \color{black} 0.000856 \color{black} \\
 \text{2,2-POS+T2} & - 2.36497 & - 0.00533 &  0.50093 &  0.50093 &  0.46511 & - 0.00039 & \color{black} 0.000314 \color{black} \\
 \text{2,3-POS} & - 2.35987 & - 0.00023 &  0.50075 &  0.50075 &  0.46547 & - 0.00003 & \color{black} 0.000029 \color{black} \\
 \text{$\approx$2,4-POS}~~ & - 2.35969 & - 0.00005 &  0.50075 &  0.50075 &  0.46549 & -0.00000 & \color{black} 0.000007 \color{black} \\
 \text{FCI} & - 2.35964 &  0.00000 &  0.50074 &  0.50074 &  0.46549 &  0.00000 & \color{black} 0.000000 \color{black}

\end{tabular}
\end{ruledtabular}

\end{table*}

{\em Results:} To illustrate the theory, we treat the molecule H$_{8}$ of eight hydrogen atoms equally space in a ring of radius $R$~\cite{Jankowski1985}.  The hydrogen atoms are represented in the minimal Slater-type-orbital (STO-3G) basis set~\cite{Hehre1969}.   The semidefinite programs are solved by the boundary-point algorithm presented in Ref.~\cite{Mazziotti2011}.  We use the Quantum Chemistry Package in Maple~\cite{QCT2022} to compute electron integrals, second-order many-body perturbation (MP2) theory, and coupled cluster with single-double excitations (CCSD) and with perturbative triples [CCSD(T)]~\cite{Bartlett2007}.  While MP2, CCSD, and CCSD(T) results are given to provide context for the difficulty in treating H$_{8}$, accurate results have been previously obtained with coupled cluster through quadruple excitations~\cite{Piecuch1994}.

The 2-RDM method is applied to computing the ground-state energy as well as several properties of the 1- and 2-RDMs for H$_{8}$ with $R=1/\sqrt{2}$~\AA\ in Table~I.  Calculations using different $N$-representability constraints are ordered with respect to their accuracy relative to full configuration interaction (FCI). Total energies reveal that the 2-RDM method converges rapidly with respect to the addition of positivity conditions. For example, the $(2,2)$-positivity, $(2,3)$-positivity, and partial $(2,4)$-positivity conditions on the 2-RDM, denoted 2,2-POS, 2,3-POS, and $\approx$2,4-POS, have energy errors of $-0.03207$, $-0.00023$, and $-0.00005$~a.u. \textcolor{black}{ and 2-RDM errors, measured relative to FCI by the $l2$ norm, of $0.001523$, $0.000029$, and $0.000007$, respectively, showing that all two-particle expectation values have an accuracy similar to the energy.}  While the $(2,q)$-positivity conditions scale in floating-point operations as $r^{3q}$, these scalings can be substantially reduced through low-rank tensor approximations; for example, exploiting the low-rank structure of the T2 constraint reduces its scaling from $r^{9}$ to $r^{6}$~\cite{Mazziotti2020} \textcolor{black}{and exploiting the structure of the $(2,4)$-positivity constraints reduces their scaling from $r^{12}$ to $r^{7}$}.  The partial $(2,4)$-positivity conditions include 2-RDM constraints derived from the D4, E4, F4, and Q4 metric matrices (not previously implemented for molecules) that physically constrain the probability distributions of four particles, three particles and a hole, three holes and a particle, and four holes~\cite{Mazziotti2001c}.  The hierarchy of constraints on the 2-RDM, generated by Eq.~(\ref{eq:B2}),  rapidly improves upon the simplest sum-of-squares conditions---the T1 or T2 condition where T1 is $^{3} D + ^{3} Q \succeq 0$ and T2 is $^{3} E+^{3} F \succeq 0$~\cite{Zhao2004, Mazziotti2005, Mazziotti2016, Mazziotti2020, Erdahl1978}.

Occupations of the highest occupied natural orbital (HONO) and the lowest unoccupied natural orbital (LUNO) are also shown in Table I.  These orbitals are half-filled from all 2-RDM calculations, regardless of the level of $N$-representability conditions, revealing the diradical character of H$_{8}$.  MP2 does not show any radical character with HONO and LUNO occupations of 0.99477 and 0.00733 while coupled cluster with single-double excitations (CCSD) partially captures the diradical character.  Results are also presented for $\langle 1/{r_{12}} \rangle$, which show that the hierarchy also converges rapidly towards other two-electron observables computable from the 2-RDM.


\begin{figure}[htp!]

\includegraphics[scale=0.4]{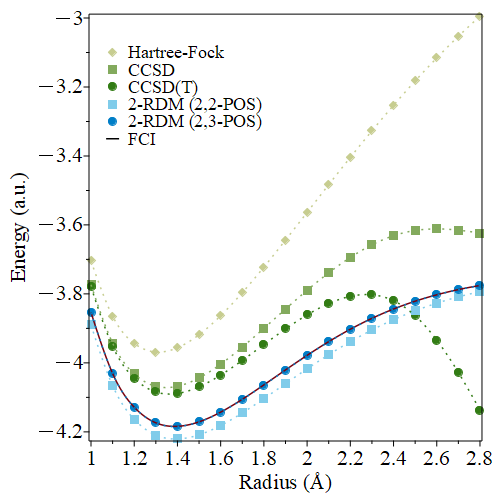}

\caption{For the H$_{8}$ ring the potential energy curves with respect to the radius $R$ from several methods including the 2-RDM calculations with $(2,2)$-positivity (2,2-POS) and $(2,3)$-positivity (2,3-POS) conditions as well as wave function calculations from the Hartree-Fock (HF), CCSD, and CCSD with perturbative triple excitations [CCSD(T)] are compared.}

\label{f:n2}

\end{figure}

Expanding the H$_{8}$ ring introduces significant static correlation~\cite{Kais2007, BenavidesRiveros2017} beyond the diradical character present at short radii.  Figure~1 compares the potential energy curves from 2-RDM calculations with 2,2-POS and 2,3-POS conditions as well as wave function calculations from Hartree-Fock (HF), CCSD, and CCSD with perturbative triple excitations [CCSD(T)].   The 2,3-POS energies significantly improve upon the 2,2-POS energies with the errors from 2,3-POS being consistently less than 0.001~a.u.  Moreover, the curves from 2,2-POS and 2,3-POS are more accurate than those from CSSD or CCSD(T).   Similarly, Fig.~2 shows that the 2-RDM methods also yield accurate expectations of $1/r_{12}$ with the errors from the 2,3-POS constraints lying between $5 \times 10^{-5}$ and $5 \times 10^{-6}$~a.u.


\begin{figure}[htp!]

\includegraphics[scale=0.4]{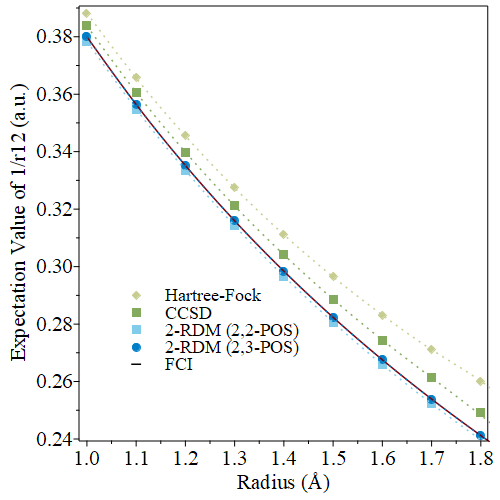}

\caption{For the H$_{8}$ ring, curves of expectation values of $1/r_{12}$ with respect to the radius $R$ from several methods including the 2-RDM calculations with (2,2)-positivity (2,2-POS) and (2,3)-positivity (2,3-POS) conditions as well as wave function calculations from the Hartree-Fock (HF), CCSD, and CCSD with perturbative triple excitations [CCSD(T)] are compared.}

\label{f:r12}

\end{figure}

{\em Discussion and Conclusions:} A complete characterization of the $N$-representability conditions is known to be NP-complete (nondeterministic polynomial-time complete) on a classical computer~\cite{Mazziotti2012b} and QMA-complete (quantum-Merlin-Arthur complete) on a quantum computer~\cite{Liu2007}---meaning that the $N$-representability conditions likely cannot be characterized in polynomial time.   The present theory is consistent with these complexity arguments in that the worst-case scenario requires an $N$-particle metric matrix, whose dimension scales exponentially with $N$.  Practically, however, we find in this work (as well as in earlier work where the $p$-positivity conditions were imposed on the $p$-RDM~\cite{Mazziotti2001c, Mazziotti2006b, Gidofalvi2006, Schwerdtfeger2009, Sinitskiy2010, Li2021}) that the energy and 2-RDM generally converge rapidly with respect to the level $p$ of the $(2,p)$-positivity conditions although there are scenarios such as the critical point of the Ising model~\cite{Schwerdtfeger2009} or spin-frustrated lattices~\cite{Sinitskiy2010, Li2021} where large $p$ may be required.  In contrast to the higher-order RDMs required in previous $p$-positivity calculations~\cite{Mazziotti2001c, Mazziotti2006b, Gidofalvi2006, Schwerdtfeger2009, Sinitskiy2010, Li2021}, the $p$-particle matrices in the polar cone have a low-rank structure which can be exploited to reduce the computational cost, as discussed in Refs.~\cite{Mazziotti2016, Mazziotti2020}.  Although the present study considers only spin and spatial symmetries, future work will explore the low-rank structure.

In summary, we present a many-body theory based on a solution of the $N$-representability problem of the 2-RDM. We parameterize the $N$-representability of the 2-RDM through a hierarchy of efficiently solvable semidefinite programs.  The variational 2-RDM calculations yield rapidly convergent lower bounds on the ground-state energy that complement the upper bounds of variational wave functions. The techniques presented here can be readily extended to other quantum marginal problems~\cite{Schilling2014}.  Application of the hierarchy to octahedral H$_{8}$ shows that it exhibits accurate energies and 2-RDMs even in the presence of strong correlation.  The solution of the $N$-representability problem---once ranked by the National Research Council as a top outstanding problem in chemical physics---provides an important alternative many-body theory based on the direct determination of the 2-RDM with important applications in the molecular sciences.

\begin{acknowledgments}

D.A.M. gratefully acknowledges the U.S. National Science Foundation Grant No. CHE-2155082 and Grant No. CHE-2035876 and the Department of Energy, Office of Basic Energy Sciences Grant DE-SC0019215.

\end{acknowledgments}

\bibliography{SFDM-22}

\begin{thebibliography}{70}%
\makeatletter
\providecommand \@ifxundefined [1]{%
 \@ifx{#1\undefined}
}%
\providecommand \@ifnum [1]{%
 \ifnum #1\expandafter \@firstoftwo
 \else \expandafter \@secondoftwo
 \fi
}%
\providecommand \@ifx [1]{%
 \ifx #1\expandafter \@firstoftwo
 \else \expandafter \@secondoftwo
 \fi
}%
\providecommand \natexlab [1]{#1}%
\providecommand \enquote  [1]{``#1''}%
\providecommand \bibnamefont  [1]{#1}%
\providecommand \bibfnamefont [1]{#1}%
\providecommand \citenamefont [1]{#1}%
\providecommand \href@noop [0]{\@secondoftwo}%
\providecommand \href [0]{\begingroup \@sanitize@url \@href}%
\providecommand \@href[1]{\@@startlink{#1}\@@href}%
\providecommand \@@href[1]{\endgroup#1\@@endlink}%
\providecommand \@sanitize@url [0]{\catcode `\\12\catcode `\$12\catcode
  `\&12\catcode `\#12\catcode `\^12\catcode `\_12\catcode `\%12\relax}%
\providecommand \@@startlink[1]{}%
\providecommand \@@endlink[0]{}%
\providecommand \url  [0]{\begingroup\@sanitize@url \@url }%
\providecommand \@url [1]{\endgroup\@href {#1}{\urlprefix }}%
\providecommand \urlprefix  [0]{URL }%
\providecommand \Eprint [0]{\href }%
\providecommand \doibase [0]{https://doi.org/}%
\providecommand \selectlanguage [0]{\@gobble}%
\providecommand \bibinfo  [0]{\@secondoftwo}%
\providecommand \bibfield  [0]{\@secondoftwo}%
\providecommand \translation [1]{[#1]}%
\providecommand \BibitemOpen [0]{}%
\providecommand \bibitemStop [0]{}%
\providecommand \bibitemNoStop [0]{.\EOS\space}%
\providecommand \EOS [0]{\spacefactor3000\relax}%
\providecommand \BibitemShut  [1]{\csname bibitem#1\endcsname}%
\let\auto@bib@innerbib\@empty
\bibitem [{\citenamefont {Mazziotti}(2007)}]{Mazziotti2007}%
  \BibitemOpen
  \bibinfo {editor} {\bibfnamefont {D.~A.}\ \bibnamefont {Mazziotti}},\ ed.,\
  \href {https://doi.org/10.1002/0470106603} {\emph {\bibinfo {title}
  {Reduced-Density-Matrix Mechanics: With Application to Many-Electron Atoms
  and Molecules}}}\ (\bibinfo  {publisher} {John Wiley {\&} Sons, Inc.},\
  \bibinfo {year} {2007})\BibitemShut {NoStop}%
\bibitem [{\citenamefont {Coleman}\ and\ \citenamefont
  {Yukalov}(2000)}]{Coleman2000}%
  \BibitemOpen
  \bibfield  {author} {\bibinfo {author} {\bibfnamefont {A.~J.}\ \bibnamefont
  {Coleman}}\ and\ \bibinfo {author} {\bibfnamefont {V.~I.}\ \bibnamefont
  {Yukalov}},\ }\href
  {https://www.ebook.de/de/product/6701945/a_j_coleman_v_i_yukalov_reduced_density_matrices.html}
  {\emph {\bibinfo {title} {Reduced Density Matrices}}}\ (\bibinfo  {publisher}
  {Springer Berlin Heidelberg},\ \bibinfo {year} {2000})\BibitemShut {NoStop}%
\bibitem [{\citenamefont {Parr}\ and\ \citenamefont {Yang}(1994)}]{DFTBook}%
  \BibitemOpen
  \bibfield  {author} {\bibinfo {author} {\bibfnamefont {R.~G.}\ \bibnamefont
  {Parr}}\ and\ \bibinfo {author} {\bibfnamefont {W.}~\bibnamefont {Yang}},\
  }\href@noop {} {\emph {\bibinfo {title} {Density-Functional Theory of Atoms
  and Molecules}}}\ (\bibinfo  {publisher} {Oxford University Press},\ \bibinfo
  {address} {New York},\ \bibinfo {year} {1994})\BibitemShut {NoStop}%
\bibitem [{\citenamefont {Medvedev}\ \emph {et~al.}(2017)\citenamefont
  {Medvedev}, \citenamefont {Bushmarinov}, \citenamefont {Sun}, \citenamefont
  {Perdew},\ and\ \citenamefont {Lyssenko}}]{Medvedev2017}%
  \BibitemOpen
  \bibfield  {author} {\bibinfo {author} {\bibfnamefont {M.~G.}\ \bibnamefont
  {Medvedev}}, \bibinfo {author} {\bibfnamefont {I.~S.}\ \bibnamefont
  {Bushmarinov}}, \bibinfo {author} {\bibfnamefont {J.}~\bibnamefont {Sun}},
  \bibinfo {author} {\bibfnamefont {J.~P.}\ \bibnamefont {Perdew}},\ and\
  \bibinfo {author} {\bibfnamefont {K.~A.}\ \bibnamefont {Lyssenko}},\
  }\bibfield  {title} {\bibinfo {title} {Density functional theory is straying
  from the path toward the exact functional},\ }\href
  {https://doi.org/10.1126/science.aah5975} {\bibfield  {journal} {\bibinfo
  {journal} {Science}\ }\textbf {\bibinfo {volume} {355}},\ \bibinfo {pages}
  {49} (\bibinfo {year} {2017})}\BibitemShut {NoStop}%
\bibitem [{\citenamefont {Coleman}(1963)}]{Coleman1963}%
  \BibitemOpen
  \bibfield  {author} {\bibinfo {author} {\bibfnamefont {A.~J.}\ \bibnamefont
  {Coleman}},\ }\bibfield  {title} {\bibinfo {title} {Structure of fermion
  density matrices},\ }\href {https://doi.org/10.1103/revmodphys.35.668}
  {\bibfield  {journal} {\bibinfo  {journal} {Rev. Mod. Phys.}\ }\textbf
  {\bibinfo {volume} {35}},\ \bibinfo {pages} {668} (\bibinfo {year}
  {1963})}\BibitemShut {NoStop}%
\bibitem [{\citenamefont {Garrod}\ and\ \citenamefont
  {Percus}(1964)}]{Garrod1964}%
  \BibitemOpen
  \bibfield  {author} {\bibinfo {author} {\bibfnamefont {C.}~\bibnamefont
  {Garrod}}\ and\ \bibinfo {author} {\bibfnamefont {J.~K.}\ \bibnamefont
  {Percus}},\ }\bibfield  {title} {\bibinfo {title} {Reduction of the
  ${N}$-particle variational problem},\ }\href
  {https://doi.org/10.1063/1.1704098} {\bibfield  {journal} {\bibinfo
  {journal} {J. Math. Phys.}\ }\textbf {\bibinfo {volume} {5}},\ \bibinfo
  {pages} {1756} (\bibinfo {year} {1964})}\BibitemShut {NoStop}%
\bibitem [{\citenamefont {Kummer}(1967)}]{Kummer1967}%
  \BibitemOpen
  \bibfield  {author} {\bibinfo {author} {\bibfnamefont {H.}~\bibnamefont
  {Kummer}},\ }\bibfield  {title} {\bibinfo {title} {${N}$-representability
  problem for reduced density matrices},\ }\href
  {https://doi.org/10.1063/1.1705122} {\bibfield  {journal} {\bibinfo
  {journal} {J. Math. Phys.}\ }\textbf {\bibinfo {volume} {8}},\ \bibinfo
  {pages} {2063} (\bibinfo {year} {1967})}\BibitemShut {NoStop}%
\bibitem [{\citenamefont {Erdahl}(1978)}]{Erdahl1978}%
  \BibitemOpen
  \bibfield  {author} {\bibinfo {author} {\bibfnamefont {R.~M.}\ \bibnamefont
  {Erdahl}},\ }\bibfield  {title} {\bibinfo {title} {Representability},\ }\href
  {https://doi.org/10.1002/qua.560130603} {\bibfield  {journal} {\bibinfo
  {journal} {Int. J. Quantum Chem.}\ }\textbf {\bibinfo {volume} {13}},\
  \bibinfo {pages} {697} (\bibinfo {year} {1978})}\BibitemShut {NoStop}%
\bibitem [{\citenamefont {Erdahl}\ and\ \citenamefont
  {Smith}(1987)}]{Erdahl1989}%
  \BibitemOpen
  \bibinfo {editor} {\bibfnamefont {R.~M.}\ \bibnamefont {Erdahl}}\ and\
  \bibinfo {editor} {\bibfnamefont {V.}~\bibnamefont {Smith}},\ eds.,\ \href
  {https://www.ebook.de/de/product/4046488/density_matrices_and_density_functionals.html}
  {\emph {\bibinfo {title} {Density Matrices and Density Functionals}}}\
  (\bibinfo  {publisher} {Springer Netherlands},\ \bibinfo {year}
  {1987})\BibitemShut {NoStop}%
\bibitem [{\citenamefont {Mazziotti}(2012)}]{Mazziotti2012b}%
  \BibitemOpen
  \bibfield  {author} {\bibinfo {author} {\bibfnamefont {D.~A.}\ \bibnamefont
  {Mazziotti}},\ }\bibfield  {title} {\bibinfo {title} {Structure of fermionic
  density matrices: Complete ${N}$-representability conditions},\ }\href
  {https://doi.org/10.1103/PhysRevLett.108.263002} {\bibfield  {journal}
  {\bibinfo  {journal} {Phys. Rev. Lett.}\ }\textbf {\bibinfo {volume} {108}},\
  \bibinfo {pages} {263002} (\bibinfo {year} {2012})}\BibitemShut {NoStop}%
\bibitem [{\citenamefont {Castillo}\ \emph {et~al.}(2021)\citenamefont
  {Castillo}, \citenamefont {Labbé}, \citenamefont {Liebert}, \citenamefont
  {Padrol}, \citenamefont {Philippe},\ and\ \citenamefont
  {Schilling}}]{Castillo2021}%
  \BibitemOpen
  \bibfield  {author} {\bibinfo {author} {\bibfnamefont {F.}~\bibnamefont
  {Castillo}}, \bibinfo {author} {\bibfnamefont {J.-P.}\ \bibnamefont
  {Labbé}}, \bibinfo {author} {\bibfnamefont {J.}~\bibnamefont {Liebert}},
  \bibinfo {author} {\bibfnamefont {A.}~\bibnamefont {Padrol}}, \bibinfo
  {author} {\bibfnamefont {E.}~\bibnamefont {Philippe}},\ and\ \bibinfo
  {author} {\bibfnamefont {C.}~\bibnamefont {Schilling}},\ }\bibfield  {title}
  {\bibinfo {title} {An effective solution to convex $1$-body
  ${N}$-representability},\ }\href@noop {} {\bibfield  {journal} {\bibinfo
  {journal} {arxiv}\ } (\bibinfo {year} {2021})},\ \Eprint
  {https://arxiv.org/abs/2105.06459} {arXiv:2105.06459 [quant-ph]} \BibitemShut
  {NoStop}%
\bibitem [{\citenamefont {Erdahl}\ and\ \citenamefont
  {Jin}(2000)}]{Erdahl2000}%
  \BibitemOpen
  \bibfield  {author} {\bibinfo {author} {\bibfnamefont {R.}~\bibnamefont
  {Erdahl}}\ and\ \bibinfo {author} {\bibfnamefont {B.}~\bibnamefont {Jin}},\
  }\bibfield  {title} {\bibinfo {title} {The lower bound method for reduced
  density matrices},\ }\href {https://doi.org/10.1016/s0166-1280(00)00494-2}
  {\bibfield  {journal} {\bibinfo  {journal} {J. Mol. Struc.}\ }\textbf
  {\bibinfo {volume} {527}},\ \bibinfo {pages} {207} (\bibinfo {year}
  {2000})}\BibitemShut {NoStop}%
\bibitem [{\citenamefont {Nakata}\ \emph {et~al.}(2001)\citenamefont {Nakata},
  \citenamefont {Nakatsuji}, \citenamefont {Ehara}, \citenamefont {Fukuda},
  \citenamefont {Nakata},\ and\ \citenamefont {Fujisawa}}]{Nakata2001}%
  \BibitemOpen
  \bibfield  {author} {\bibinfo {author} {\bibfnamefont {M.}~\bibnamefont
  {Nakata}}, \bibinfo {author} {\bibfnamefont {H.}~\bibnamefont {Nakatsuji}},
  \bibinfo {author} {\bibfnamefont {M.}~\bibnamefont {Ehara}}, \bibinfo
  {author} {\bibfnamefont {M.}~\bibnamefont {Fukuda}}, \bibinfo {author}
  {\bibfnamefont {K.}~\bibnamefont {Nakata}},\ and\ \bibinfo {author}
  {\bibfnamefont {K.}~\bibnamefont {Fujisawa}},\ }\bibfield  {title} {\bibinfo
  {title} {Variational calculations of fermion second-order reduced density
  matrices by semidefinite programming algorithm},\ }\href
  {https://doi.org/10.1063/1.1360199} {\bibfield  {journal} {\bibinfo
  {journal} {J. Chem. Phys.}\ }\textbf {\bibinfo {volume} {114}},\ \bibinfo
  {pages} {8282} (\bibinfo {year} {2001})}\BibitemShut {NoStop}%
\bibitem [{\citenamefont {Mazziotti}\ and\ \citenamefont
  {Erdahl}(2001)}]{Mazziotti2001c}%
  \BibitemOpen
  \bibfield  {author} {\bibinfo {author} {\bibfnamefont {D.~A.}\ \bibnamefont
  {Mazziotti}}\ and\ \bibinfo {author} {\bibfnamefont {R.~M.}\ \bibnamefont
  {Erdahl}},\ }\bibfield  {title} {\bibinfo {title} {Uncertainty relations and
  reduced density matrices: Mapping many-body quantum mechanics onto four
  particles},\ }\href {https://doi.org/10.1103/PhysRevA.63.042113} {\bibfield
  {journal} {\bibinfo  {journal} {Phys. Rev. A}\ }\textbf {\bibinfo {volume}
  {63}},\ \bibinfo {pages} {042113} (\bibinfo {year} {2001})}\BibitemShut
  {NoStop}%
\bibitem [{\citenamefont {Mazziotti}(2002)}]{Mazziotti2002b}%
  \BibitemOpen
  \bibfield  {author} {\bibinfo {author} {\bibfnamefont {D.~A.}\ \bibnamefont
  {Mazziotti}},\ }\bibfield  {title} {\bibinfo {title} {Variational
  minimization of atomic and molecular ground-state energies via the
  two-particle reduced density matrix},\ }\href
  {https://doi.org/10.1103/PhysRevA.65.062511} {\bibfield  {journal} {\bibinfo
  {journal} {Phys. Rev. A}\ }\textbf {\bibinfo {volume} {65}},\ \bibinfo
  {pages} {062511} (\bibinfo {year} {2002})}\BibitemShut {NoStop}%
\bibitem [{\citenamefont {Zhao}\ \emph {et~al.}(2004)\citenamefont {Zhao},
  \citenamefont {Braams}, \citenamefont {Fukuda}, \citenamefont {Overton},\
  and\ \citenamefont {Percus}}]{Zhao2004}%
  \BibitemOpen
  \bibfield  {author} {\bibinfo {author} {\bibfnamefont {Z.}~\bibnamefont
  {Zhao}}, \bibinfo {author} {\bibfnamefont {B.~J.}\ \bibnamefont {Braams}},
  \bibinfo {author} {\bibfnamefont {M.}~\bibnamefont {Fukuda}}, \bibinfo
  {author} {\bibfnamefont {M.~L.}\ \bibnamefont {Overton}},\ and\ \bibinfo
  {author} {\bibfnamefont {J.~K.}\ \bibnamefont {Percus}},\ }\bibfield  {title}
  {\bibinfo {title} {The reduced density matrix method for electronic structure
  calculations and the role of three-index representability conditions},\
  }\href {https://doi.org/10.1063/1.1636721} {\bibfield  {journal} {\bibinfo
  {journal} {J. Chem. Phys.}\ }\textbf {\bibinfo {volume} {120}},\ \bibinfo
  {pages} {2095} (\bibinfo {year} {2004})}\BibitemShut {NoStop}%
\bibitem [{\citenamefont {Mazziotti}(2004)}]{Mazziotti2004a}%
  \BibitemOpen
  \bibfield  {author} {\bibinfo {author} {\bibfnamefont {D.~A.}\ \bibnamefont
  {Mazziotti}},\ }\bibfield  {title} {\bibinfo {title} {Realization of quantum
  chemistry without wave functions through first-order semidefinite
  programming},\ }\href {https://doi.org/10.1103/PhysRevLett.93.213001}
  {\bibfield  {journal} {\bibinfo  {journal} {Phys. Rev. Lett.}\ }\textbf
  {\bibinfo {volume} {93}},\ \bibinfo {pages} {213001} (\bibinfo {year}
  {2004})}\BibitemShut {NoStop}%
\bibitem [{\citenamefont {Mazziotti}(2005)}]{Mazziotti2005}%
  \BibitemOpen
  \bibfield  {author} {\bibinfo {author} {\bibfnamefont {D.~A.}\ \bibnamefont
  {Mazziotti}},\ }\bibfield  {title} {\bibinfo {title} {Variational
  two-electron reduced density matrix theory for many-electron atoms and
  molecules: Implementation of the spin- and symmetry-adapted {T2} condition
  through first-order semidefinite programming},\ }\href
  {https://doi.org/10.1103/physreva.72.032510} {\bibfield  {journal} {\bibinfo
  {journal} {Phys. Rev. A}\ }\textbf {\bibinfo {volume} {72}},\ \bibinfo
  {pages} {032510} (\bibinfo {year} {2005})}\BibitemShut {NoStop}%
\bibitem [{\citenamefont {Canc{\`{e}}s}\ \emph {et~al.}(2006)\citenamefont
  {Canc{\`{e}}s}, \citenamefont {Stoltz},\ and\ \citenamefont
  {Lewin}}]{Cances2006}%
  \BibitemOpen
  \bibfield  {author} {\bibinfo {author} {\bibfnamefont {E.}~\bibnamefont
  {Canc{\`{e}}s}}, \bibinfo {author} {\bibfnamefont {G.}~\bibnamefont
  {Stoltz}},\ and\ \bibinfo {author} {\bibfnamefont {M.}~\bibnamefont
  {Lewin}},\ }\bibfield  {title} {\bibinfo {title} {The electronic ground-state
  energy problem: A new reduced density matrix approach},\ }\href
  {https://doi.org/10.1063/1.2222358} {\bibfield  {journal} {\bibinfo
  {journal} {J. Chem. Phys.}\ }\textbf {\bibinfo {volume} {125}},\ \bibinfo
  {pages} {064101} (\bibinfo {year} {2006})}\BibitemShut {NoStop}%
\bibitem [{\citenamefont {Mazziotti}(2006)}]{Mazziotti2006b}%
  \BibitemOpen
  \bibfield  {author} {\bibinfo {author} {\bibfnamefont {D.~A.}\ \bibnamefont
  {Mazziotti}},\ }\bibfield  {title} {\bibinfo {title} {Variational
  reduced-density-matrix method using three-particle ${N}$-representability
  conditions with application to many-electron molecules},\ }\href
  {https://doi.org/10.1103/PhysRevA.74.032501} {\bibfield  {journal} {\bibinfo
  {journal} {Phys. Rev. A}\ }\textbf {\bibinfo {volume} {74}},\ \bibinfo
  {pages} {032501} (\bibinfo {year} {2006})}\BibitemShut {NoStop}%
\bibitem [{\citenamefont {Erdahl}(2007)}]{Erdahl2007}%
  \BibitemOpen
  \bibfield  {author} {\bibinfo {author} {\bibfnamefont {R.~M.}\ \bibnamefont
  {Erdahl}},\ }\bibfield  {title} {\bibinfo {title} {The lower bound method for
  density matrices and semidefinite programming},\ }in\ \href
  {https://doi.org/10.1002/9780470106600.ch4} {\emph {\bibinfo {booktitle}
  {Reduced-Density-Matrix Mechanics: With Application to Many-Electron Atoms
  and Molecules}}}\ (\bibinfo  {publisher} {John Wiley {\&} Sons, Inc.},\
  \bibinfo {year} {2007})\ pp.\ \bibinfo {pages} {61--91}\BibitemShut {NoStop}%
\bibitem [{\citenamefont {Braams}\ \emph {et~al.}(2007)\citenamefont {Braams},
  \citenamefont {Percus},\ and\ \citenamefont {Zhao}}]{Braams2007}%
  \BibitemOpen
  \bibfield  {author} {\bibinfo {author} {\bibfnamefont {B.~J.}\ \bibnamefont
  {Braams}}, \bibinfo {author} {\bibfnamefont {J.~K.}\ \bibnamefont {Percus}},\
  and\ \bibinfo {author} {\bibfnamefont {Z.}~\bibnamefont {Zhao}},\ }\bibfield
  {title} {\bibinfo {title} {The {T1} and {T2} representability conditions},\
  }in\ \href {https://doi.org/10.1002/9780470106600.ch5} {\emph {\bibinfo
  {booktitle} {Reduced-Density-Matrix Mechanics: With Application to
  Many-Electron Atoms and Molecules}}}\ (\bibinfo  {publisher} {John Wiley {\&}
  Sons, Inc.},\ \bibinfo {year} {2007})\ pp.\ \bibinfo {pages}
  {93--101}\BibitemShut {NoStop}%
\bibitem [{\citenamefont {Gidofalvi}\ and\ \citenamefont
  {Mazziotti}(2008)}]{Gidofalvi2008}%
  \BibitemOpen
  \bibfield  {author} {\bibinfo {author} {\bibfnamefont {G.}~\bibnamefont
  {Gidofalvi}}\ and\ \bibinfo {author} {\bibfnamefont {D.~A.}\ \bibnamefont
  {Mazziotti}},\ }\bibfield  {title} {\bibinfo {title} {Active-space
  two-electron reduced-density-matrix method: Complete active-space
  calculations without diagonalization of the ${N}$-electron {H}amiltonian},\
  }\href {https://doi.org/10.1063/1.2983652} {\bibfield  {journal} {\bibinfo
  {journal} {J. Chem. Phys.}\ }\textbf {\bibinfo {volume} {129}},\ \bibinfo
  {pages} {134108} (\bibinfo {year} {2008})}\BibitemShut {NoStop}%
\bibitem [{\citenamefont {Shenvi}\ and\ \citenamefont
  {Izmaylov}(2010)}]{Shenvi2010}%
  \BibitemOpen
  \bibfield  {author} {\bibinfo {author} {\bibfnamefont {N.}~\bibnamefont
  {Shenvi}}\ and\ \bibinfo {author} {\bibfnamefont {A.~F.}\ \bibnamefont
  {Izmaylov}},\ }\bibfield  {title} {\bibinfo {title} {Active-{Space}
  ${N}$-representability constraints for variational two-particle reduced
  density matrix calculations},\ }\href
  {https://doi.org/10.1103/physrevlett.105.213003} {\bibfield  {journal}
  {\bibinfo  {journal} {Phys. Rev. Lett.}\ }\textbf {\bibinfo {volume} {105}},\
  \bibinfo {pages} {213003} (\bibinfo {year} {2010})}\BibitemShut {NoStop}%
\bibitem [{\citenamefont {Mazziotti}(2011)}]{Mazziotti2011}%
  \BibitemOpen
  \bibfield  {author} {\bibinfo {author} {\bibfnamefont {D.~A.}\ \bibnamefont
  {Mazziotti}},\ }\bibfield  {title} {\bibinfo {title} {Large-scale
  semidefinite programming for many-electron quantum mechanics},\ }\href
  {https://doi.org/10.1103/PhysRevLett.106.083001} {\bibfield  {journal}
  {\bibinfo  {journal} {Phys. Rev. Lett.}\ }\textbf {\bibinfo {volume} {106}},\
  \bibinfo {pages} {083001} (\bibinfo {year} {2011})}\BibitemShut {NoStop}%
\bibitem [{\citenamefont {Verstichel}\ \emph {et~al.}(2012)\citenamefont
  {Verstichel}, \citenamefont {van Aggelen}, \citenamefont {Poelmans},\ and\
  \citenamefont {{Van Neck}}}]{Verstichel2012}%
  \BibitemOpen
  \bibfield  {author} {\bibinfo {author} {\bibfnamefont {B.}~\bibnamefont
  {Verstichel}}, \bibinfo {author} {\bibfnamefont {H.}~\bibnamefont {van
  Aggelen}}, \bibinfo {author} {\bibfnamefont {W.}~\bibnamefont {Poelmans}},\
  and\ \bibinfo {author} {\bibfnamefont {D.}~\bibnamefont {{Van Neck}}},\
  }\bibfield  {title} {\bibinfo {title} {Variational two-particle density
  matrix calculation for the {H}ubbard model below half filling using
  spin-adapted lifting conditions},\ }\href
  {https://doi.org/10.1103/physrevlett.108.213001} {\bibfield  {journal}
  {\bibinfo  {journal} {Phys. Rev. Lett.}\ }\textbf {\bibinfo {volume} {108}},\
  \bibinfo {pages} {213001} (\bibinfo {year} {2012})}\BibitemShut {NoStop}%
\bibitem [{\citenamefont {Baumgratz}\ and\ \citenamefont
  {Plenio}(2012)}]{Baumgratz2012}%
  \BibitemOpen
  \bibfield  {author} {\bibinfo {author} {\bibfnamefont {T.}~\bibnamefont
  {Baumgratz}}\ and\ \bibinfo {author} {\bibfnamefont {M.~B.}\ \bibnamefont
  {Plenio}},\ }\bibfield  {title} {\bibinfo {title} {Lower bounds for ground
  states of condensed matter systems},\ }\href
  {https://doi.org/10.1088/1367-2630/14/2/023027} {\bibfield  {journal}
  {\bibinfo  {journal} {New J. Phys.}\ }\textbf {\bibinfo {volume} {14}},\
  \bibinfo {pages} {023027} (\bibinfo {year} {2012})}\BibitemShut {NoStop}%
\bibitem [{\citenamefont {Poelmans}\ \emph {et~al.}(2015)\citenamefont
  {Poelmans}, \citenamefont {Raemdonck}, \citenamefont {Verstichel},
  \citenamefont {Baerdemacker}, \citenamefont {Torre}, \citenamefont {Lain},
  \citenamefont {Massaccesi}, \citenamefont {Alcoba}, \citenamefont
  {Bultinck},\ and\ \citenamefont {{Van Neck}}}]{Poelmans2015}%
  \BibitemOpen
  \bibfield  {author} {\bibinfo {author} {\bibfnamefont {W.}~\bibnamefont
  {Poelmans}}, \bibinfo {author} {\bibfnamefont {M.~V.}\ \bibnamefont
  {Raemdonck}}, \bibinfo {author} {\bibfnamefont {B.}~\bibnamefont
  {Verstichel}}, \bibinfo {author} {\bibfnamefont {S.~D.}\ \bibnamefont
  {Baerdemacker}}, \bibinfo {author} {\bibfnamefont {A.}~\bibnamefont {Torre}},
  \bibinfo {author} {\bibfnamefont {L.}~\bibnamefont {Lain}}, \bibinfo {author}
  {\bibfnamefont {G.~E.}\ \bibnamefont {Massaccesi}}, \bibinfo {author}
  {\bibfnamefont {D.~R.}\ \bibnamefont {Alcoba}}, \bibinfo {author}
  {\bibfnamefont {P.}~\bibnamefont {Bultinck}},\ and\ \bibinfo {author}
  {\bibfnamefont {D.}~\bibnamefont {{Van Neck}}},\ }\bibfield  {title}
  {\bibinfo {title} {Variational optimization of the second-order density
  matrix corresponding to a seniority-zero configuration interaction wave
  function},\ }\href {https://doi.org/10.1021/acs.jctc.5b00378} {\bibfield
  {journal} {\bibinfo  {journal} {J. Chem. Theory Comput.}\ }\textbf {\bibinfo
  {volume} {11}},\ \bibinfo {pages} {4064} (\bibinfo {year}
  {2015})}\BibitemShut {NoStop}%
\bibitem [{\citenamefont {Mazziotti}(2016)}]{Mazziotti2016}%
  \BibitemOpen
  \bibfield  {author} {\bibinfo {author} {\bibfnamefont {D.~A.}\ \bibnamefont
  {Mazziotti}},\ }\bibfield  {title} {\bibinfo {title} {Enhanced constraints
  for accurate lower bounds on many-electron quantum energies from variational
  two-electron reduced density matrix theory},\ }\href
  {https://doi.org/10.1103/PhysRevLett.117.153001} {\bibfield  {journal}
  {\bibinfo  {journal} {Phys. Rev. Lett.}\ }\textbf {\bibinfo {volume} {117}},\
  \bibinfo {pages} {153001} (\bibinfo {year} {2016})}\BibitemShut {NoStop}%
\bibitem [{\citenamefont {Alcoba}\ \emph {et~al.}(2018)\citenamefont {Alcoba},
  \citenamefont {Torre}, \citenamefont {Lain}, \citenamefont {Massaccesi},
  \citenamefont {O{\~{n}}a}, \citenamefont {Honor{\'{e}}}, \citenamefont
  {Poelmans}, \citenamefont {{Van Neck}}, \citenamefont {Bultinck},\ and\
  \citenamefont {Baerdemacker}}]{Alcoba2018}%
  \BibitemOpen
  \bibfield  {author} {\bibinfo {author} {\bibfnamefont {D.~R.}\ \bibnamefont
  {Alcoba}}, \bibinfo {author} {\bibfnamefont {A.}~\bibnamefont {Torre}},
  \bibinfo {author} {\bibfnamefont {L.}~\bibnamefont {Lain}}, \bibinfo {author}
  {\bibfnamefont {G.~E.}\ \bibnamefont {Massaccesi}}, \bibinfo {author}
  {\bibfnamefont {O.~B.}\ \bibnamefont {O{\~{n}}a}}, \bibinfo {author}
  {\bibfnamefont {E.~M.}\ \bibnamefont {Honor{\'{e}}}}, \bibinfo {author}
  {\bibfnamefont {W.}~\bibnamefont {Poelmans}}, \bibinfo {author}
  {\bibfnamefont {D.}~\bibnamefont {{Van Neck}}}, \bibinfo {author}
  {\bibfnamefont {P.}~\bibnamefont {Bultinck}},\ and\ \bibinfo {author}
  {\bibfnamefont {S.~D.}\ \bibnamefont {Baerdemacker}},\ }\bibfield  {title}
  {\bibinfo {title} {Direct variational determination of the two-electron
  reduced density matrix for doubly occupied-configuration-interaction wave
  functions: The influence of three-index ${N}$-representability conditions},\
  }\href {https://doi.org/10.1063/1.5008811} {\bibfield  {journal} {\bibinfo
  {journal} {J. Chem. Phys.}\ }\textbf {\bibinfo {volume} {148}},\ \bibinfo
  {pages} {024105} (\bibinfo {year} {2018})}\BibitemShut {NoStop}%
\bibitem [{\citenamefont {Rubio-Garc{\'{\i}}a}\ \emph
  {et~al.}(2019)\citenamefont {Rubio-Garc{\'{\i}}a}, \citenamefont {Dukelsky},
  \citenamefont {Alcoba}, \citenamefont {Capuzzi}, \citenamefont {O{\~{n}}a},
  \citenamefont {R{\'{\i}}os}, \citenamefont {Torre},\ and\ \citenamefont
  {Lain}}]{Rubio-Garcia2019}%
  \BibitemOpen
  \bibfield  {author} {\bibinfo {author} {\bibfnamefont {A.}~\bibnamefont
  {Rubio-Garc{\'{\i}}a}}, \bibinfo {author} {\bibfnamefont {J.}~\bibnamefont
  {Dukelsky}}, \bibinfo {author} {\bibfnamefont {D.~R.}\ \bibnamefont
  {Alcoba}}, \bibinfo {author} {\bibfnamefont {P.}~\bibnamefont {Capuzzi}},
  \bibinfo {author} {\bibfnamefont {O.~B.}\ \bibnamefont {O{\~{n}}a}}, \bibinfo
  {author} {\bibfnamefont {E.}~\bibnamefont {R{\'{\i}}os}}, \bibinfo {author}
  {\bibfnamefont {A.}~\bibnamefont {Torre}},\ and\ \bibinfo {author}
  {\bibfnamefont {L.}~\bibnamefont {Lain}},\ }\bibfield  {title} {\bibinfo
  {title} {Variational reduced density matrix method in the doubly-occupied
  configuration interaction space using four-particle ${N}$-representability
  conditions: Application to the {XXZ} model of quantum magnetism},\ }\href
  {https://doi.org/10.1063/1.5118899} {\bibfield  {journal} {\bibinfo
  {journal} {J. Chem. Phys.}\ }\textbf {\bibinfo {volume} {151}},\ \bibinfo
  {pages} {154104} (\bibinfo {year} {2019})}\BibitemShut {NoStop}%
\bibitem [{\citenamefont {Head-Marsden}\ and\ \citenamefont
  {Mazziotti}(2020)}]{Head-Marsden2020}%
  \BibitemOpen
  \bibfield  {author} {\bibinfo {author} {\bibfnamefont {K.}~\bibnamefont
  {Head-Marsden}}\ and\ \bibinfo {author} {\bibfnamefont {D.~A.}\ \bibnamefont
  {Mazziotti}},\ }\bibfield  {title} {\bibinfo {title} {Active-space pair
  two-electron reduced density matrix theory for strong correlation},\ }\href
  {https://doi.org/10.1021/acs.jpca.0c01937} {\bibfield  {journal} {\bibinfo
  {journal} {J. Phys. Chem. A}\ }\textbf {\bibinfo {volume} {124}},\ \bibinfo
  {pages} {4848} (\bibinfo {year} {2020})}\BibitemShut {NoStop}%
\bibitem [{\citenamefont {Haim}\ \emph {et~al.}()\citenamefont {Haim},
  \citenamefont {Kueng},\ and\ \citenamefont {Refael}}]{Haim2020}%
  \BibitemOpen
  \bibfield  {author} {\bibinfo {author} {\bibfnamefont {A.}~\bibnamefont
  {Haim}}, \bibinfo {author} {\bibfnamefont {R.}~\bibnamefont {Kueng}},\ and\
  \bibinfo {author} {\bibfnamefont {G.}~\bibnamefont {Refael}},\ }\bibfield
  {title} {\bibinfo {title} {Variational-correlations approach to quantum
  many-body problems},\ }\href@noop {} {\bibfield  {journal} {\bibinfo
  {journal} {arxiv}\ }}\Eprint
  {https://arxiv.org/abs/http://arxiv.org/abs/2001.06510v1}
  {http://arxiv.org/abs/2001.06510v1} \BibitemShut {NoStop}%
\bibitem [{\citenamefont {Han}()}]{Han2020}%
  \BibitemOpen
  \bibfield  {author} {\bibinfo {author} {\bibfnamefont {X.}~\bibnamefont
  {Han}},\ }\bibfield  {title} {\bibinfo {title} {Quantum many-body
  bootstrap},\ }\href@noop {} {\bibfield  {journal} {\bibinfo  {journal}
  {arxiv}\ }}\Eprint {https://arxiv.org/abs/http://arxiv.org/abs/2006.06002v1}
  {http://arxiv.org/abs/2006.06002v1} \BibitemShut {NoStop}%
\bibitem [{\citenamefont {Mazziotti}(2020)}]{Mazziotti2020}%
  \BibitemOpen
  \bibfield  {author} {\bibinfo {author} {\bibfnamefont {D.~A.}\ \bibnamefont
  {Mazziotti}},\ }\bibfield  {title} {\bibinfo {title} {Dual-cone variational
  calculation of the two-electron reduced density matrix},\ }\href
  {https://doi.org/10.1103/physreva.102.052819} {\bibfield  {journal} {\bibinfo
   {journal} {Phys. Rev. A}\ }\textbf {\bibinfo {volume} {102}},\ \bibinfo
  {pages} {052819} (\bibinfo {year} {2020})}\BibitemShut {NoStop}%
\bibitem [{\citenamefont {Li}\ \emph {et~al.}(2021)\citenamefont {Li},
  \citenamefont {Liebenthal},\ and\ \citenamefont {DePrince}}]{Li2021}%
  \BibitemOpen
  \bibfield  {author} {\bibinfo {author} {\bibfnamefont {R.~R.}\ \bibnamefont
  {Li}}, \bibinfo {author} {\bibfnamefont {M.~D.}\ \bibnamefont {Liebenthal}},\
  and\ \bibinfo {author} {\bibfnamefont {A.~E.}\ \bibnamefont {DePrince}},\
  }\bibfield  {title} {\bibinfo {title} {Challenges for variational
  reduced-density-matrix theory with three-particle ${N}$-representability
  conditions},\ }\href {https://doi.org/10.1063/5.0066404} {\bibfield
  {journal} {\bibinfo  {journal} {J. Chem. Phys.}\ }\textbf {\bibinfo {volume}
  {155}},\ \bibinfo {pages} {174110} (\bibinfo {year} {2021})}\BibitemShut
  {NoStop}%
\bibitem [{\citenamefont {Knight}\ \emph {et~al.}(2022)\citenamefont {Knight},
  \citenamefont {Quiney},\ and\ \citenamefont {Martin}}]{Knight2022}%
  \BibitemOpen
  \bibfield  {author} {\bibinfo {author} {\bibfnamefont {M.~J.}\ \bibnamefont
  {Knight}}, \bibinfo {author} {\bibfnamefont {H.~M.}\ \bibnamefont {Quiney}},\
  and\ \bibinfo {author} {\bibfnamefont {A.~M.}\ \bibnamefont {Martin}},\
  }\bibfield  {title} {\bibinfo {title} {Reduced density matrix approach to
  ultracold few-fermion systems in one dimension},\ }\href
  {https://doi.org/10.1088/1367-2630/ac643d} {\bibfield  {journal} {\bibinfo
  {journal} {New J. Phys.}\ }\textbf {\bibinfo {volume} {24}},\ \bibinfo
  {pages} {053004} (\bibinfo {year} {2022})}\BibitemShut {NoStop}%
\bibitem [{\citenamefont {Xie}\ \emph {et~al.}(2022)\citenamefont {Xie},
  \citenamefont {Ewing}, \citenamefont {Boyn}, \citenamefont {Filatov},
  \citenamefont {Cheng}, \citenamefont {Ma}, \citenamefont {Grocke},
  \citenamefont {Zhao}, \citenamefont {Itani}, \citenamefont {Sun},
  \citenamefont {Cho}, \citenamefont {Chen}, \citenamefont {Chapman},
  \citenamefont {Patel}, \citenamefont {Talapin}, \citenamefont {Park},
  \citenamefont {Mazziotti},\ and\ \citenamefont {Anderson}}]{Xie2022}%
  \BibitemOpen
  \bibfield  {author} {\bibinfo {author} {\bibfnamefont {J.}~\bibnamefont
  {Xie}}, \bibinfo {author} {\bibfnamefont {S.}~\bibnamefont {Ewing}}, \bibinfo
  {author} {\bibfnamefont {J.-N.}\ \bibnamefont {Boyn}}, \bibinfo {author}
  {\bibfnamefont {A.~S.}\ \bibnamefont {Filatov}}, \bibinfo {author}
  {\bibfnamefont {B.}~\bibnamefont {Cheng}}, \bibinfo {author} {\bibfnamefont
  {T.}~\bibnamefont {Ma}}, \bibinfo {author} {\bibfnamefont {G.~L.}\
  \bibnamefont {Grocke}}, \bibinfo {author} {\bibfnamefont {N.}~\bibnamefont
  {Zhao}}, \bibinfo {author} {\bibfnamefont {R.}~\bibnamefont {Itani}},
  \bibinfo {author} {\bibfnamefont {X.}~\bibnamefont {Sun}}, \bibinfo {author}
  {\bibfnamefont {H.}~\bibnamefont {Cho}}, \bibinfo {author} {\bibfnamefont
  {Z.}~\bibnamefont {Chen}}, \bibinfo {author} {\bibfnamefont {K.~W.}\
  \bibnamefont {Chapman}}, \bibinfo {author} {\bibfnamefont {S.~N.}\
  \bibnamefont {Patel}}, \bibinfo {author} {\bibfnamefont {D.~V.}\ \bibnamefont
  {Talapin}}, \bibinfo {author} {\bibfnamefont {J.}~\bibnamefont {Park}},
  \bibinfo {author} {\bibfnamefont {D.~A.}\ \bibnamefont {Mazziotti}},\ and\
  \bibinfo {author} {\bibfnamefont {J.~S.}\ \bibnamefont {Anderson}},\
  }\bibfield  {title} {\bibinfo {title} {Intrinsic glassy-metallic transport in
  an amorphous coordination polymer},\ }\href
  {https://doi.org/10.1038/s41586-022-05261-4} {\bibfield  {journal} {\bibinfo
  {journal} {Nature}\ }\textbf {\bibinfo {volume} {611}},\ \bibinfo {pages}
  {479} (\bibinfo {year} {2022})}\BibitemShut {NoStop}%
\bibitem [{\citenamefont {Safaei}\ and\ \citenamefont
  {Mazziotti}(2018)}]{Safaei2018}%
  \BibitemOpen
  \bibfield  {author} {\bibinfo {author} {\bibfnamefont {S.}~\bibnamefont
  {Safaei}}\ and\ \bibinfo {author} {\bibfnamefont {D.~A.}\ \bibnamefont
  {Mazziotti}},\ }\bibfield  {title} {\bibinfo {title} {Quantum signature of
  exciton condensation},\ }\href {https://doi.org/10.1103/PhysRevB.98.045122}
  {\bibfield  {journal} {\bibinfo  {journal} {Phys. Rev. B}\ }\textbf {\bibinfo
  {volume} {98}},\ \bibinfo {pages} {045122} (\bibinfo {year}
  {2018})}\BibitemShut {NoStop}%
\bibitem [{\citenamefont {Schouten}\ \emph {et~al.}(2021)\citenamefont
  {Schouten}, \citenamefont {Sager},\ and\ \citenamefont
  {Mazziotti}}]{Schouten2021}%
  \BibitemOpen
  \bibfield  {author} {\bibinfo {author} {\bibfnamefont {A.~O.}\ \bibnamefont
  {Schouten}}, \bibinfo {author} {\bibfnamefont {L.~M.}\ \bibnamefont
  {Sager}},\ and\ \bibinfo {author} {\bibfnamefont {D.~A.}\ \bibnamefont
  {Mazziotti}},\ }\bibfield  {title} {\bibinfo {title} {Exciton condensation in
  molecular-scale van der {W}aals stacks},\ }\href
  {https://doi.org/10.1021/acs.jpclett.1c02368} {\bibfield  {journal} {\bibinfo
   {journal} {J. Phys. Chem. Lett.}\ }\textbf {\bibinfo {volume} {12}},\
  \bibinfo {pages} {9906} (\bibinfo {year} {2021})}\BibitemShut {NoStop}%
\bibitem [{\citenamefont {Schouten}\ \emph {et~al.}(2022)\citenamefont
  {Schouten}, \citenamefont {Sager-Smith},\ and\ \citenamefont
  {Mazziotti}}]{Schouten2022}%
  \BibitemOpen
  \bibfield  {author} {\bibinfo {author} {\bibfnamefont {A.~O.}\ \bibnamefont
  {Schouten}}, \bibinfo {author} {\bibfnamefont {L.~M.}\ \bibnamefont
  {Sager-Smith}},\ and\ \bibinfo {author} {\bibfnamefont {D.~A.}\ \bibnamefont
  {Mazziotti}},\ }\bibfield  {title} {\bibinfo {title} {Large cumulant
  eigenvalue as a signature of exciton condensation},\ }\href
  {https://doi.org/10.1103/physrevb.105.245151} {\bibfield  {journal} {\bibinfo
   {journal} {Phys. Rev. B}\ }\textbf {\bibinfo {volume} {105}},\ \bibinfo
  {pages} {245151} (\bibinfo {year} {2022})}\BibitemShut {NoStop}%
\bibitem [{\citenamefont {Sager}\ \emph {et~al.}(2022)\citenamefont {Sager},
  \citenamefont {Schouten},\ and\ \citenamefont {Mazziotti}}]{Sager2022}%
  \BibitemOpen
  \bibfield  {author} {\bibinfo {author} {\bibfnamefont {L.~M.}\ \bibnamefont
  {Sager}}, \bibinfo {author} {\bibfnamefont {A.~O.}\ \bibnamefont
  {Schouten}},\ and\ \bibinfo {author} {\bibfnamefont {D.~A.}\ \bibnamefont
  {Mazziotti}},\ }\bibfield  {title} {\bibinfo {title} {Beginnings of exciton
  condensation in coronene analog of graphene double layer},\ }\href
  {https://doi.org/10.1063/5.0084564} {\bibfield  {journal} {\bibinfo
  {journal} {J. Chem. Phys.}\ }\textbf {\bibinfo {volume} {156}},\ \bibinfo
  {pages} {154702} (\bibinfo {year} {2022})}\BibitemShut {NoStop}%
\bibitem [{\citenamefont {Schilling}\ and\ \citenamefont
  {Pittalis}(2021)}]{Schilling2021}%
  \BibitemOpen
  \bibfield  {author} {\bibinfo {author} {\bibfnamefont {C.}~\bibnamefont
  {Schilling}}\ and\ \bibinfo {author} {\bibfnamefont {S.}~\bibnamefont
  {Pittalis}},\ }\bibfield  {title} {\bibinfo {title} {Ensemble reduced density
  matrix functional theory for excited states and hierarchical generalization
  of pauli's exclusion principle},\ }\href
  {https://doi.org/10.1103/physrevlett.127.023001} {\bibfield  {journal}
  {\bibinfo  {journal} {Phys. Rev. Lett.}\ }\textbf {\bibinfo {volume} {127}},\
  \bibinfo {pages} {023001} (\bibinfo {year} {2021})}\BibitemShut {NoStop}%
\bibitem [{\citenamefont {Benavides-Riveros}\ \emph {et~al.}(2020)\citenamefont
  {Benavides-Riveros}, \citenamefont {Wolff}, \citenamefont {Marques},\ and\
  \citenamefont {Schilling}}]{Benavides-Riveros2020}%
  \BibitemOpen
  \bibfield  {author} {\bibinfo {author} {\bibfnamefont {C.~L.}\ \bibnamefont
  {Benavides-Riveros}}, \bibinfo {author} {\bibfnamefont {J.}~\bibnamefont
  {Wolff}}, \bibinfo {author} {\bibfnamefont {M.~A.}\ \bibnamefont {Marques}},\
  and\ \bibinfo {author} {\bibfnamefont {C.}~\bibnamefont {Schilling}},\
  }\bibfield  {title} {\bibinfo {title} {Reduced density matrix functional
  theory for bosons},\ }\href {https://doi.org/10.1103/physrevlett.124.180603}
  {\bibfield  {journal} {\bibinfo  {journal} {Phys. Rev. Lett.}\ }\textbf
  {\bibinfo {volume} {124}},\ \bibinfo {pages} {180603} (\bibinfo {year}
  {2020})}\BibitemShut {NoStop}%
\bibitem [{\citenamefont {Schilling}\ and\ \citenamefont
  {Schilling}(2019)}]{Schilling2019}%
  \BibitemOpen
  \bibfield  {author} {\bibinfo {author} {\bibfnamefont {C.}~\bibnamefont
  {Schilling}}\ and\ \bibinfo {author} {\bibfnamefont {R.}~\bibnamefont
  {Schilling}},\ }\bibfield  {title} {\bibinfo {title} {Diverging exchange
  force and form of the exact density matrix functional},\ }\href
  {https://doi.org/10.1103/PhysRevLett.122.013001} {\bibfield  {journal}
  {\bibinfo  {journal} {Phys. Rev. Lett.}\ }\textbf {\bibinfo {volume} {122}},\
  \bibinfo {pages} {013001} (\bibinfo {year} {2019})}\BibitemShut {NoStop}%
\bibitem [{\citenamefont {Piris}(2021)}]{Piris2021}%
  \BibitemOpen
  \bibfield  {author} {\bibinfo {author} {\bibfnamefont {M.}~\bibnamefont
  {Piris}},\ }\bibfield  {title} {\bibinfo {title} {Global natural orbital
  functional: Towards the complete description of the electron correlation},\
  }\href {https://doi.org/10.1103/physrevlett.127.233001} {\bibfield  {journal}
  {\bibinfo  {journal} {Phys. Rev. Lett.}\ }\textbf {\bibinfo {volume} {127}},\
  \bibinfo {pages} {233001} (\bibinfo {year} {2021})}\BibitemShut {NoStop}%
\bibitem [{\citenamefont {Piris}(2017)}]{Piris2017a}%
  \BibitemOpen
  \bibfield  {author} {\bibinfo {author} {\bibfnamefont {M.}~\bibnamefont
  {Piris}},\ }\bibfield  {title} {\bibinfo {title} {Global method for electron
  correlation},\ }\href {https://doi.org/10.1103/PhysRevLett.119.063002}
  {\bibfield  {journal} {\bibinfo  {journal} {Phys. Rev. Lett.}\ }\textbf
  {\bibinfo {volume} {119}},\ \bibinfo {pages} {063002} (\bibinfo {year}
  {2017})}\BibitemShut {NoStop}%
\bibitem [{\citenamefont {Gibney}\ \emph {et~al.}(2022)\citenamefont {Gibney},
  \citenamefont {Boyn},\ and\ \citenamefont {Mazziotti}}]{Gibney2022a}%
  \BibitemOpen
  \bibfield  {author} {\bibinfo {author} {\bibfnamefont {D.}~\bibnamefont
  {Gibney}}, \bibinfo {author} {\bibfnamefont {J.-N.}\ \bibnamefont {Boyn}},\
  and\ \bibinfo {author} {\bibfnamefont {D.~A.}\ \bibnamefont {Mazziotti}},\
  }\bibfield  {title} {\bibinfo {title} {Density functional theory transformed
  into a one-electron reduced-density-matrix functional theory for the capture
  of static correlation},\ }\href {https://doi.org/10.1021/acs.jpclett.2c00083}
  {\bibfield  {journal} {\bibinfo  {journal} {J. Phys. Chem. Lett.}\ }\textbf
  {\bibinfo {volume} {13}},\ \bibinfo {pages} {1382} (\bibinfo {year}
  {2022})}\BibitemShut {NoStop}%
\bibitem [{\citenamefont {Coleman}\ and\ \citenamefont
  {Absar}(1980)}]{Coleman1980}%
  \BibitemOpen
  \bibfield  {author} {\bibinfo {author} {\bibfnamefont {A.~J.}\ \bibnamefont
  {Coleman}}\ and\ \bibinfo {author} {\bibfnamefont {I.}~\bibnamefont
  {Absar}},\ }\bibfield  {title} {\bibinfo {title} {Reduced hamiltonian
  orbitals. {III}. {U}nitarily invariant decomposition of hermitian
  operators},\ }\href {https://doi.org/10.1002/qua.560180513} {\bibfield
  {journal} {\bibinfo  {journal} {Int. J. Quantum Chem.}\ }\textbf {\bibinfo
  {volume} {18}},\ \bibinfo {pages} {1279} (\bibinfo {year}
  {1980})}\BibitemShut {NoStop}%
\bibitem [{\citenamefont {Au-Chin}\ and\ \citenamefont
  {Hong}(1983)}]{AuChin1983}%
  \BibitemOpen
  \bibfield  {author} {\bibinfo {author} {\bibfnamefont {T.}~\bibnamefont
  {Au-Chin}}\ and\ \bibinfo {author} {\bibfnamefont {G.}~\bibnamefont {Hong}},\
  }\bibfield  {title} {\bibinfo {title} {Characteristic operators and unitarily
  invariant decomposition of hermitian operators},\ }\href
  {https://doi.org/10.1002/qua.560230120} {\bibfield  {journal} {\bibinfo
  {journal} {Int. J. Quantum Chem.}\ }\textbf {\bibinfo {volume} {23}},\
  \bibinfo {pages} {217} (\bibinfo {year} {1983})}\BibitemShut {NoStop}%
\bibitem [{\citenamefont {Vandenberghe}\ and\ \citenamefont
  {Boyd}(1996)}]{Vandenberghe1996}%
  \BibitemOpen
  \bibfield  {author} {\bibinfo {author} {\bibfnamefont {L.}~\bibnamefont
  {Vandenberghe}}\ and\ \bibinfo {author} {\bibfnamefont {S.}~\bibnamefont
  {Boyd}},\ }\bibfield  {title} {\bibinfo {title} {Semidefinite programming},\
  }\href {https://doi.org/10.1137/1038003} {\bibfield  {journal} {\bibinfo
  {journal} {{SIAM} Review}\ }\textbf {\bibinfo {volume} {38}},\ \bibinfo
  {pages} {49} (\bibinfo {year} {1996})}\BibitemShut {NoStop}%
\bibitem [{\citenamefont {Fukuda}\ \emph {et~al.}(2007)\citenamefont {Fukuda},
  \citenamefont {Nakata},\ and\ \citenamefont {Yamashita}}]{Fukuda2007}%
  \BibitemOpen
  \bibfield  {author} {\bibinfo {author} {\bibfnamefont {M.}~\bibnamefont
  {Fukuda}}, \bibinfo {author} {\bibfnamefont {M.}~\bibnamefont {Nakata}},\
  and\ \bibinfo {author} {\bibfnamefont {M.}~\bibnamefont {Yamashita}},\
  }\bibfield  {title} {\bibinfo {title} {Semidefinite programming: Formulations
  and primal-dual interior-point methods},\ }in\ \href
  {https://doi.org/10.1002/9780470106600.ch6} {\emph {\bibinfo {booktitle}
  {Reduced-Density-Matrix Mechanics: With Application to Many-Electron Atoms
  and Molecules}}}\ (\bibinfo  {publisher} {John Wiley {\&} Sons, Inc.},\
  \bibinfo {year} {2007})\ pp.\ \bibinfo {pages} {103--118}\BibitemShut
  {NoStop}%
\bibitem [{\citenamefont {Erdahl}(1979)}]{Erdahl1979}%
  \BibitemOpen
  \bibfield  {author} {\bibinfo {author} {\bibfnamefont {R.}~\bibnamefont
  {Erdahl}},\ }\bibfield  {title} {\bibinfo {title} {Two algorithms for the
  lower bound method of reduced density matrix theory},\ }\href
  {https://doi.org/10.1016/0034-4877(79)90015-6} {\bibfield  {journal}
  {\bibinfo  {journal} {Rep. Math. Phys.}\ }\textbf {\bibinfo {volume} {15}},\
  \bibinfo {pages} {147} (\bibinfo {year} {1979})}\BibitemShut {NoStop}%
\bibitem [{\citenamefont {Gidofalvi}\ and\ \citenamefont
  {Mazziotti}(2006)}]{Gidofalvi2006}%
  \BibitemOpen
  \bibfield  {author} {\bibinfo {author} {\bibfnamefont {G.}~\bibnamefont
  {Gidofalvi}}\ and\ \bibinfo {author} {\bibfnamefont {D.~A.}\ \bibnamefont
  {Mazziotti}},\ }\bibfield  {title} {\bibinfo {title} {Computation of quantum
  phase transitions by reduced-density-matrix mechanics},\ }\href
  {https://doi.org/10.1103/physreva.74.012501} {\bibfield  {journal} {\bibinfo
  {journal} {Phys. Rev. A}\ }\textbf {\bibinfo {volume} {74}},\ \bibinfo
  {pages} {012501} (\bibinfo {year} {2006})}\BibitemShut {NoStop}%
\bibitem [{\citenamefont {Schwerdtfeger}\ and\ \citenamefont
  {Mazziotti}(2009)}]{Schwerdtfeger2009}%
  \BibitemOpen
  \bibfield  {author} {\bibinfo {author} {\bibfnamefont {C.~A.}\ \bibnamefont
  {Schwerdtfeger}}\ and\ \bibinfo {author} {\bibfnamefont {D.~A.}\ \bibnamefont
  {Mazziotti}},\ }\bibfield  {title} {\bibinfo {title} {Convex-set description
  of quantum phase transitions in the transverse {I}sing model using
  reduced-density-matrix theory},\ }\href {https://doi.org/10.1063/1.3143403}
  {\bibfield  {journal} {\bibinfo  {journal} {J. Chem. Phys.}\ }\textbf
  {\bibinfo {volume} {130}},\ \bibinfo {pages} {224102} (\bibinfo {year}
  {2009})}\BibitemShut {NoStop}%
\bibitem [{\citenamefont {Gidofalvi}\ and\ \citenamefont
  {Mazziotti}(2004)}]{Gidofalvi2004}%
  \BibitemOpen
  \bibfield  {author} {\bibinfo {author} {\bibfnamefont {G.}~\bibnamefont
  {Gidofalvi}}\ and\ \bibinfo {author} {\bibfnamefont {D.~A.}\ \bibnamefont
  {Mazziotti}},\ }\bibfield  {title} {\bibinfo {title} {Boson correlation
  energies via variational minimization with the two-particle reduced density
  matrix: Exact ${N}$-representability conditions for harmonic interactions},\
  }\href {https://doi.org/10.1103/physreva.69.042511} {\bibfield  {journal}
  {\bibinfo  {journal} {Phys. Rev. A}\ }\textbf {\bibinfo {volume} {69}},\
  \bibinfo {pages} {042511} (\bibinfo {year} {2004})}\BibitemShut {NoStop}%
\bibitem [{\citenamefont {Slebodzi\'nski}(1970)}]{Slebodzinski1970}%
  \BibitemOpen
  \bibfield  {author} {\bibinfo {author} {\bibfnamefont {W.}~\bibnamefont
  {Slebodzi\'nski}},\ }\href@noop {} {\emph {\bibinfo {title} {Exterior Forms
  and their Applications}}}\ (\bibinfo  {publisher} {Polish Scientific
  Publishers, Warsaw},\ \bibinfo {year} {1970})\BibitemShut {NoStop}%
\bibitem [{\citenamefont {Mazziotti}(1998)}]{Mazziotti1998}%
  \BibitemOpen
  \bibfield  {author} {\bibinfo {author} {\bibfnamefont {D.~A.}\ \bibnamefont
  {Mazziotti}},\ }\bibfield  {title} {\bibinfo {title} {Contracted
  {S}chrödinger equation: {D}etermining quantum energies and two-particle
  density matrices without wave functions},\ }\href
  {https://doi.org/10.1103/physreva.57.4219} {\bibfield  {journal} {\bibinfo
  {journal} {Phys. Rev. A}\ }\textbf {\bibinfo {volume} {57}},\ \bibinfo
  {pages} {4219} (\bibinfo {year} {1998})}\BibitemShut {NoStop}%
\bibitem [{\citenamefont {Sinitskiy}\ \emph {et~al.}(2010)\citenamefont
  {Sinitskiy}, \citenamefont {Greenman},\ and\ \citenamefont
  {Mazziotti}}]{Sinitskiy2010}%
  \BibitemOpen
  \bibfield  {author} {\bibinfo {author} {\bibfnamefont {A.~V.}\ \bibnamefont
  {Sinitskiy}}, \bibinfo {author} {\bibfnamefont {L.}~\bibnamefont
  {Greenman}},\ and\ \bibinfo {author} {\bibfnamefont {D.~A.}\ \bibnamefont
  {Mazziotti}},\ }\bibfield  {title} {\bibinfo {title} {Strong correlation in
  hydrogen chains and lattices using the variational two-electron reduced
  density matrix method},\ }\href {https://doi.org/10.1063/1.3459059}
  {\bibfield  {journal} {\bibinfo  {journal} {J. Chem. Phys.}\ }\textbf
  {\bibinfo {volume} {133}},\ \bibinfo {pages} {014104} (\bibinfo {year}
  {2010})}\BibitemShut {NoStop}%
\bibitem [{Note1()}]{Note1}%
  \BibitemOpen
  \bibinfo {note} {See Supplemental Material at URL for $^{3} B_{i,j}$ matrices
  of the 3-positivity conditions.}\BibitemShut {Stop}%
\bibitem [{\citenamefont {Feynman}(1939)}]{Feynman1939}%
  \BibitemOpen
  \bibfield  {author} {\bibinfo {author} {\bibfnamefont {R.~P.}\ \bibnamefont
  {Feynman}},\ }\bibfield  {title} {\bibinfo {title} {Forces in molecules},\
  }\href {https://doi.org/10.1103/physrev.56.340} {\bibfield  {journal}
  {\bibinfo  {journal} {Phys. Rev.}\ }\textbf {\bibinfo {volume} {56}},\
  \bibinfo {pages} {340} (\bibinfo {year} {1939})}\BibitemShut {NoStop}%
\bibitem [{\citenamefont {Jankowski}\ \emph {et~al.}(1985)\citenamefont
  {Jankowski}, \citenamefont {Meissner},\ and\ \citenamefont
  {Wasilewski}}]{Jankowski1985}%
  \BibitemOpen
  \bibfield  {author} {\bibinfo {author} {\bibfnamefont {K.}~\bibnamefont
  {Jankowski}}, \bibinfo {author} {\bibfnamefont {L.}~\bibnamefont
  {Meissner}},\ and\ \bibinfo {author} {\bibfnamefont {J.}~\bibnamefont
  {Wasilewski}},\ }\bibfield  {title} {\bibinfo {title} {Davidson-type
  corrections for quasidegenerate states},\ }\href
  {https://doi.org/10.1002/qua.560280622} {\bibfield  {journal} {\bibinfo
  {journal} {Int. J. Quantum Chem.}\ }\textbf {\bibinfo {volume} {28}},\
  \bibinfo {pages} {931} (\bibinfo {year} {1985})}\BibitemShut {NoStop}%
\bibitem [{\citenamefont {Hehre}\ \emph {et~al.}(1969)\citenamefont {Hehre},
  \citenamefont {Stewart},\ and\ \citenamefont {Pople}}]{Hehre1969}%
  \BibitemOpen
  \bibfield  {author} {\bibinfo {author} {\bibfnamefont {W.~J.}\ \bibnamefont
  {Hehre}}, \bibinfo {author} {\bibfnamefont {R.~F.}\ \bibnamefont {Stewart}},\
  and\ \bibinfo {author} {\bibfnamefont {J.~A.}\ \bibnamefont {Pople}},\
  }\bibfield  {title} {\bibinfo {title} {Self-consistent molecular-orbital
  methods. i. {U}se of {G}aussian expansions of {S}later-type atomic
  orbitals},\ }\href {https://doi.org/10.1063/1.1672392} {\bibfield  {journal}
  {\bibinfo  {journal} {J. Chem. Phys.}\ }\textbf {\bibinfo {volume} {51}},\
  \bibinfo {pages} {2657} (\bibinfo {year} {1969})}\BibitemShut {NoStop}%
\bibitem [{\citenamefont {RDMChem}(2022)}]{QCT2022}%
  \BibitemOpen
  \bibfield  {author} {\bibinfo {author} {\bibnamefont {RDMChem}},\ }\href@noop
  {} {\emph {\bibinfo {title} {{Q}uantum {C}hemistry {P}ackage in {M}aple}}}\
  (\bibinfo  {publisher} {Maplesoft, Waterloo},\ \bibinfo {year}
  {2022})\BibitemShut {NoStop}%
\bibitem [{\citenamefont {Bartlett}\ and\ \citenamefont
  {Musia{\l}}(2007)}]{Bartlett2007}%
  \BibitemOpen
  \bibfield  {author} {\bibinfo {author} {\bibfnamefont {R.~J.}\ \bibnamefont
  {Bartlett}}\ and\ \bibinfo {author} {\bibfnamefont {M.}~\bibnamefont
  {Musia{\l}}},\ }\bibfield  {title} {\bibinfo {title} {Coupled-cluster theory
  in quantum chemistry},\ }\href {https://doi.org/10.1103/revmodphys.79.291}
  {\bibfield  {journal} {\bibinfo  {journal} {Rev. Mod. Phys.}\ }\textbf
  {\bibinfo {volume} {79}},\ \bibinfo {pages} {291} (\bibinfo {year}
  {2007})}\BibitemShut {NoStop}%
\bibitem [{\citenamefont {Piecuch}\ and\ \citenamefont
  {Adamowicz}(1994)}]{Piecuch1994}%
  \BibitemOpen
  \bibfield  {author} {\bibinfo {author} {\bibfnamefont {P.}~\bibnamefont
  {Piecuch}}\ and\ \bibinfo {author} {\bibfnamefont {L.}~\bibnamefont
  {Adamowicz}},\ }\bibfield  {title} {\bibinfo {title} {State-selective
  multireference coupled-cluster theory employing the single-reference
  formalism: Implementation and application to the {H}$_{8}$ model system},\
  }\href {https://doi.org/10.1063/1.467143} {\bibfield  {journal} {\bibinfo
  {journal} {J. Chem. Phys.}\ }\textbf {\bibinfo {volume} {100}},\ \bibinfo
  {pages} {5792} (\bibinfo {year} {1994})}\BibitemShut {NoStop}%
\bibitem [{\citenamefont {Kais}(2007)}]{Kais2007}%
  \BibitemOpen
  \bibfield  {author} {\bibinfo {author} {\bibfnamefont {S.}~\bibnamefont
  {Kais}},\ }\bibfield  {title} {\bibinfo {title} {Entanglement, electron
  correlation, and density matrices},\ }in\ \href
  {https://doi.org/10.1002/9780470106600.ch18} {\emph {\bibinfo {booktitle}
  {Reduced-Density-Matrix Mechanics: With Application to Many-Electron Atoms
  and Molecules}}}\ (\bibinfo  {publisher} {John Wiley {\&} Sons, Inc.},\
  \bibinfo {year} {2007})\ pp.\ \bibinfo {pages} {493--535}\BibitemShut
  {NoStop}%
\bibitem [{\citenamefont {Benavides-Riveros}\ \emph {et~al.}(2017)\citenamefont
  {Benavides-Riveros}, \citenamefont {Lathiotakis},\ and\ \citenamefont
  {Marques}}]{BenavidesRiveros2017}%
  \BibitemOpen
  \bibfield  {author} {\bibinfo {author} {\bibfnamefont {C.~L.}\ \bibnamefont
  {Benavides-Riveros}}, \bibinfo {author} {\bibfnamefont {N.~N.}\ \bibnamefont
  {Lathiotakis}},\ and\ \bibinfo {author} {\bibfnamefont {M.~A.~L.}\
  \bibnamefont {Marques}},\ }\bibfield  {title} {\bibinfo {title} {Towards a
  formal definition of static and dynamic electronic correlations},\ }\href
  {https://doi.org/10.1039/c7cp01137g} {\bibfield  {journal} {\bibinfo
  {journal} {Phys. Chem. Chem. Phys.}\ }\textbf {\bibinfo {volume} {19}},\
  \bibinfo {pages} {12655} (\bibinfo {year} {2017})}\BibitemShut {NoStop}%
\bibitem [{\citenamefont {Liu}\ \emph {et~al.}(2007)\citenamefont {Liu},
  \citenamefont {Christandl},\ and\ \citenamefont {Verstraete}}]{Liu2007}%
  \BibitemOpen
  \bibfield  {author} {\bibinfo {author} {\bibfnamefont {Y.-K.}\ \bibnamefont
  {Liu}}, \bibinfo {author} {\bibfnamefont {M.}~\bibnamefont {Christandl}},\
  and\ \bibinfo {author} {\bibfnamefont {F.}~\bibnamefont {Verstraete}},\
  }\bibfield  {title} {\bibinfo {title} {Quantum computational complexity of
  the ${N}$-representability problem: {QMA} complete},\ }\href
  {https://doi.org/10.1103/physrevlett.98.110503} {\bibfield  {journal}
  {\bibinfo  {journal} {Physical Review Letters}\ }\textbf {\bibinfo {volume}
  {98}},\ \bibinfo {pages} {110503} (\bibinfo {year} {2007})}\BibitemShut
  {NoStop}%
\bibitem [{\citenamefont {Schilling}(2014)}]{Schilling2014}%
  \BibitemOpen
  \bibfield  {author} {\bibinfo {author} {\bibfnamefont {C.}~\bibnamefont
  {Schilling}},\ }\bibfield  {title} {\bibinfo {title} {The quantum marginal
  problem},\ }in\ \href {https://doi.org/10.1142/9789814618144_0010} {\emph
  {\bibinfo {booktitle} {Mathematical Results in Quantum Mechanics}}}\
  (\bibinfo  {publisher} {World Scientific},\ \bibinfo {year}
  {2014})\BibitemShut {NoStop}%
\end{thebibliography}%

\end{document}